\documentclass[journal]{IEEEtran}
\usepackage{epsfig}
\usepackage{amsmath}
\def\be{\begin{equation}}
\def\ee{\end{equation}}
\def\e#1{\label{#1}\end{equation}}
\def\bea{\begin{eqnarray}}
\def\eea{\end{eqnarray}}
\def\ea#1{\label{#1}\end{eqnarray}}

\def\bx{{\mbox{\boldmath $x$}}}
\def\bn{{\mbox{\boldmath $n$}}}

\def\br{{\mbox{\boldmath $r$}}}

\def\bsig{{\mbox{\boldmath $\sigma$}}}
\begin{document}
\title{Quantum interference in resonant tunneling and single 
spin measurements}
\author{Shmuel A. Gurvitz%
  \thanks{Based on work presented at the 2004 IEEE NTC Quantum Device
    Technology Workshop.}%
\thanks{S.A. Gurvitz is with the Department of 
Particle Physics, Weizmann Institute of Science, Rehovot 76100, 
Israel.}}
\markboth{IEEE TRANSACTIONS ON NANOTECHNOLOGY,~Vol.~4, 
No.~1,~JANUARY~2005}{Shell \MakeLowercase{\textit{et al.}}: 
                                Bare Demo of IEEEtran.cls for
                                Journals}
\maketitle
\begin{abstract}
  We consider the resonant tunneling through a multi-level
  system. It is demonstrated that the resonant
  current displays quantum interference effects due to
  a possibility of tunneling through different
  levels. We show that the 
  interference effects are strongly modulated by a relative phase of
  states carrying the current. This makes it possible to
  use these effects for measuring the phase difference between 
  resonant states in quantum dots. We extend our model for a description
  of magnetotransport through the Zeeman doublets. It is shown
  that, due to spin-flip transitions, the quantum interference effects 
  generate a distinct peak in the shot-noise power spectrum 
  at the frequency of Zeeman splitting. 
  This mechanism explains modulation in the tunneling current at the Larmor frequency
  observed in scanning tunneling microscope experiments
  and can be utilized for a single spin measurement.
   
\end{abstract}
\IEEEpeerreviewmaketitle

\section{Introduction}

\PARstart{T}{he resonant} tunneling through quantum dots (or 
impurities) has been investigated both theoretically and experimentally
in large amount of works, yet most investigations concentrated
on the resonant tunneling through a single quantum level. In the
case of the resonant tunneling through many levels one usually 
considered the total current as
a sum of currents through individual levels. In general, however, this procedure 
cannot be correct due to the quantum interference effects.
We illustrate this point with a simple example.  

\begin{figure}[htb]
\centering
  \epsfig{file=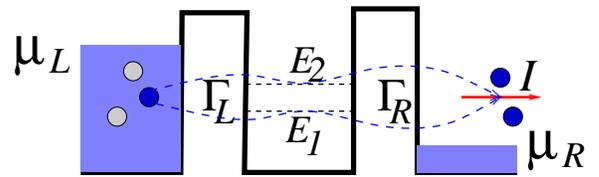,width=3.in}
 \protect\caption{
   Interference effect in the resonant tunneling through two
   levels. Here $\Gamma_{L,R}$ denotes the partial width of each of the
   levels due to tunneling to the left or right
   reservoir.
}
\label{f1}
\end{figure}

Let us consider the resonant tunneling through a quantum dot 
coupled with two reservoirs with different chemical
potentials, $\mu_{L,R}$. We assume that two levels of the dot,
$E_{1,2}$, are inside the potential bias $\mu_L-\mu_R$
(see Fig.~\ref{f1}). Then the electric current flows from the left (emitter) to
the right (collector) reservoirs through the two levels. If we neglect
the Coulomb repulsion between the electrons,  
the total current is given by the Landauer formula
\begin{equation}
  I={e\over 2\pi}\int T(E)\, dE\, ,
 \label{a0}
\end{equation} 
where $T(E)$ is the total transmission. 
Since any electron from the left reservoir
can tunnel to the right reservoirs via these two levels
(Fig.~\ref{f1}), the total transmission 
is given by a sum of two Breight-Wigner amplitudes:
\begin{equation}
T(E)= \left |{\Gamma\over E-E_1+i\Gamma}
+\eta {\Gamma\over E-E_2+i\Gamma}\right |^2\, ,
\label{a1}
\end{equation}
where $\Gamma =(\Gamma_L+\Gamma_R)/2 $ is a half of the total width.
We assume that $\Gamma_L =\Gamma_R$ (a symmetric dot), and that the
tunneling widths are the same for both levels, yet 
the Breight-Wigner amplitudes can differ in phase. We therefore
introduced the factor $\eta$ in (\ref{a1})
which denotes the relative phase of these amplitudes. It can be shown\cite{hac} 
that $\eta$ can take only two values $\pm 1$, the so-called ``in'' or
``out-of-phase'' resonances, respectively.  

It follows from (\ref{a1}) that, if the resonances do not overlap,
$\Gamma\ll E_2-E_1$, the total resonant current is a
sum of the resonant currents flowing through the levels $E_1$ and
$E_2$. However, if $\Gamma\sim E_2-E_1$, the
interference plays a very important role in the total resonant
current. Indeed, one easily finds that, in the case of constructive 
interference, $\eta =1$, the total current increases with $\Gamma$ as
$I\sim \Gamma$ and, in the case of destructive
interference, $\eta=-1$, the total current
decreases with $\Gamma$ as $1/\Gamma$. Since in the case of quantum
dots the tunneling widths $\Gamma_{L,R}$ can be varied by the
corresponding gate voltage, one can use this interference effect in order
to measure the relative phase of different levels. This can
provide an alternative method for a measurement of this quantity, 
in addition to that which utilized the Aharonov-Bohm oscillations\cite{yac}.

The interference effects described above are related to the 
stationary current. We can also anticipate the interference
effects in temporal characteristics of the current. Indeed, it is
known that the average resonant current trough a {\em double-dot
system} would display damped oscillations generated by
quantum interference\cite{gur3}. Since any double-well potential 
can be mapped to a single well with two levels,
one can expect to observe similar damped oscillations
in the resonant current flowing through two levels of a single dot. 
The presence of oscillations
in the average current is usually reflected in the current shot-noise
power spectrum density, $S(\omega )$. For instance, the resonant
current through a double-dot structure
would develop a dip in the $S(\omega )$ at the Rabi
frequency\cite{bran}. Similarly, one can think that the
current flowing through two levels (Fig.\ref{f1}) would develop
the same structure in $S(\omega )$ at $\omega =E_2-E_1$.
However, the result should strongly depend on the relative phase of
two levels. This phase-dependence of the spectral density has not been
discussed in the literature, although this effect can have important 
applications.

The interference effects in resonant tunneling can also be anticipated 
in the magnetotransport\cite{moz}.  
Indeed, in the presence of magnetic field, all levels in quantum dot
or impurity are split (Zeeman splitting). Therefore, electrons in the left reservoir
with different orientation of spin (parallel
or anti-parallel to the magnetic field) would tunnel to the right
reservoir through the
different Zeeman sublevels of the dot. This alone cannot 
result in quantum interference, since the corresponding spin states  
are orthogonal. However, if the $g$-factor in the dot is different from
that in the reservoirs, then the spin-orbit interaction generates the
spin-flip of an electron traveling through the quantum dot\cite{moz}. As a
result, the same electron from the left reservoir flows to the
right reservoir through two Zeeman sublevels 
(Fig.\ref{f1}). Similar to the previous case, one can expect 
that the related interference effects would be reflected in the behavior of the current
spectral density $S(\omega )$. In particular, it was argued 
in\cite{moz} that, in this way, one can explain the puzzled oscillations at 
Larmour frequency observed in  scanning tunneling microscope (STM)
experiments\cite{menas,dur} and considered as a promising tool for a
single spin measurement\cite{mano,dur1}.

In this paper, we investigate the interference effects in resonant
tunneling through multilevel systems as quantum dots or impurities.
At first sight, the treatment of these effects looks rather
straightforward in terms of single electron description
[(\ref{a0}) and (\ref{a1})]. However, this is not the
case when the electron-electron repulsion inside the dot is taken
into account. In fact, this effect can never be disregarded, and it always
plays a very important role in the electron transport. 
For this reason, one uses 
the Keldysh nonequilibrium Green's function technique\cite{kel,bus}
for an account of the interaction effect in the electron transport.
These calculations, however, are rather complicated and are usually performed only
in a weak coupling limit. In this paper, we use a different, simpler,
and more transparent technique developed by us in Ref.~\cite{gp}-\cite{gur2}
that consists of reduction of the Schr\"odinger equation to Bloch-type rate equations
for the density matrix obtained by integrating over the reservoir states. 
Such a procedure can be carried out
in the strong nonequilibrium limit without any stochastic assumptions 
and valid beyond the weak coupling limit. The resulting equations can
be used straightforwardly for evaluating the current in a
multilevel system and its power spectrum, with the Coulomb repulsion inside
the dots taken into account.  

The remainder of this paper is as following. In Section II we study the
resonant tunneling through two levels of the quantum dot. We obtain
the generalized quantum rate equations describing the entire system,
including the electric current. 
Special attention is paid to effect of the relative phase of
resonances on the average current and on the shot-noise power spectrum.   
In Section III, we concentrate on the magnetotransport through
quantum dots or impurities. We derive the rate equations for this
case and evaluate the shot-noise power spectrum. The obtained results
suggest a natural explanation of a peak at the Larmor density
and the hyperfine splitting due to
interaction with nuclear spin found in new  STM measurements.
This provides us with a possibility of a single nuclear spin
detection. Section IV provides a summary.

\section{Resonant tunneling through different levels}

Let us consider resonant transport in a multilevel system. 
We shall treat this problem in the framework of a tunnel
Hamiltonian approach. Therefore, we introduce the following tunneling Hamiltonian
describing the electron transport from the emitter to the
collector via different levels of a quantum dot (impurity) $E_j$
(Fig.~1), $H=H_L+H_R+H_D+H_T$, with 
\begin{eqnarray}
{}&&H_{L(R)}=\sum_{l(r)}E_{l(r)}a^\dagger_{l(r)}a_{l(r)}\, ,
  ~
  H_D=\sum_jE_jd^\dagger_jd_j+\hat U_C\, ,
  \nonumber\\
&&H_T=\Big (\sum_{l,j}\Omega^{(j)}_ld^\dagger_ja_l
+l\leftrightarrow r\Big )+H.c.\, .
  \label{a2}
\end{eqnarray}
Here $a^\dagger_{l,r}(a_{l,r})$ is the creation (annihilation)
operator of an electron in the reservoirs and $d^\dagger_j(d_j)$
is the same operator for an electron in the dot (we omitted
the spin indices). The operator
$\hat U_C=\sum_{jj'}(U_C/2)d^\dagger_jd_jd^\dagger_{j'}d_{j'}$
denotes the Coulomb interaction of between electrons in the dot and 
$\Omega^{(j)}_l(\Omega^{(j)}_r)$ is a coupling between 
the states $E_l(E_r)$ and $E_j$ of the reservoir and the dot,
respectively. This coupling is related to the corresponding
tunneling width by
$\Gamma^{(j)}_{L,R}=2\pi\rho_{L,R}|\Omega^{(j)}_{l,r}|^2$,
where $\rho_{L,R}$ is the density of states in the corresponding
reservoir. (In the absence of magnetic field, one can always choose
the gauge such that all couplings $\Omega$ are real).

All parameters of the tunneling Hamiltonian (\ref{a2})
are related to the initial microscopic description of the system
in the configuration space ($\bx $). 
For instance, the coupling $\Omega^{(j)}_{l(r)}$
is given by the Bardeen formula\cite{bar}
\begin{equation} 
\Omega^{(j)}_{l(r)}=-{1\over 2m}\int_{\bx\in\Sigma_{l(r)}}\phi_j(\bx )
\stackrel{\leftrightarrow}\nabla_{\bn}\chi_{l(r)}(\bx )d\sigma\ ,  
\label{a3}
\end{equation}
where $\phi_j(\bx )$ and $\chi_{l(r)}(\bx )$ are the
electron wave functions inside the dot and the reservoir,
respectively, and $\Sigma$ is a surface inside the potential barrier
that separates the dot from the corresponding
reservoir. In one-dimensional (1-D) case
$\phi_j(\bx )\equiv \phi_j(x)$ and $\chi_{l(r)}(\bx )\equiv \chi_{l(r)}(x)$,
(\ref{a3}) can be rewritten as\cite{g1}
\begin{equation} 
\Omega^{(j)}_{l(r)}=-(\kappa_j/m)\phi_j(\bar x_{l(r)})\chi_{l(r)}
(\bar x_{l(r)})\, ,
\label{a4}
\end{equation}
where $\kappa_j=\sqrt{2m(V(\bar x)-E_j)}$. The point
$\bar x_{l(r)}$ should be taken inside the left (right)
barrier and far away from the classical
turning points where $\Omega^{(j)}_{l(r)}$ becomes practically independent
of $\bar x$\cite{g2}.

It was demonstrated in \cite{gp}-\cite{gur1} that the Schr\"odinger equation
$i\partial_t|\Psi (t)\rangle =H|\Psi (t)\rangle$, describing the 
quantum transport through a multidot system,
can be transformed to the  Bloch-type rate equation for
the reduced density-matrix
$\sigma_{\alpha\beta}^n(t)\equiv \sigma_{\alpha\beta}^{nn}(t)$, where 
$|\alpha\rangle,\,  |\beta\rangle,\ldots$ are the discrete states of 
the system in the occupation number representation and $n$ is the
number of electrons arriving at the corresponding reservoir by time $t$.
This reduction takes place after partial tracing
over the reservoir states, and it becomes the exact one in the limit of large bias
$\mu_L-\mu_R\gg \Gamma_{L,R}$ without explicit use of any Markov-type or weak coupling
approximations. As a result, the off-diagonal in $n$ 
density matrix elements, $\sigma_{\alpha\beta}^{nn'}(t)$, 
becomes decoupled from the diagonal in $n$ terms,
$\sigma_{\alpha\beta}^{n}(t)$, in the equations of motion\cite{gur2}.
Finally, one arrives at the following Bloch-type
equations describing the entire system \cite{gur1}: 
\begin{eqnarray}
&&\dot\sigma_{\alpha\beta}^n=i\epsilon_{\beta\alpha}\sigma_{\alpha\beta}^n +
i\left (\sum_{\gamma}\sigma_{\alpha\gamma}^n
\tilde\Omega_{\gamma\to\beta}
-\sum_{\gamma}\tilde\Omega_{\alpha\to\gamma}
\sigma_{\gamma\beta}^n\right )\nonumber\\
&&-{{\sum}_{\gamma,\delta}}
\pi\rho(\sigma_{\alpha\gamma}^n\Omega_{\gamma\to\delta}\Omega_{\delta\to\beta}
+\sigma_{\gamma\beta}^n\Omega_{\gamma\to\delta}\Omega_{\delta\to\alpha})
\nonumber\\
&&+\sum_{\gamma,\delta}
\pi\rho\,(\Omega_{\gamma\to\alpha}\Omega_{\delta\to\beta}+
\Omega_{\gamma\to\beta}\Omega_{\delta\to\alpha})
\sigma_{\gamma\delta}^{n-1}\, ,
\label{a5}
\end{eqnarray}
where $\epsilon_{\beta\alpha}=E_\beta -E_\alpha$ and 
$\Omega_{\alpha\to\beta}$ denotes one-electron 
hopping amplitude that generates $\alpha\to\beta$ transition.
We distinguish between the amplitudes $\tilde\Omega$ and $\Omega$ of   
one-electron hopping among isolated states and among isolated
and continuum states, respectively. The latter transitions are 
of the second order in the hopping amplitude $\sim\Omega^2$. 
These transition are produced by two consecutive hoppings of an
electron across continuum states with the density of states
$\rho$.  

Solving (\ref{a5}), we can determine the probability of finding
$n$ electrons in the collector, $P_n(t)=\sum_j\sigma_{jj}^n(t)$.
This quantity allows us to determine the average current
\begin{equation} 
I(t)=e\sum_nn\dot P(t)\, ,
\label{a6}
\end{equation}
and the current power spectrum. The latter is given by the McDonald
formula\cite{mac,moz}
\begin{equation}
S(\omega) = 2e^2\omega \int_0^\infty dt
\sin (\omega t) {d\over dt}N_R^2 (t)\, ,
\label{a7}
\end{equation}
where $N_R^2 (t) =\sum_n n^2P_n(t)$.

\begin{figure}[htb]
\centering
  \epsfig{file=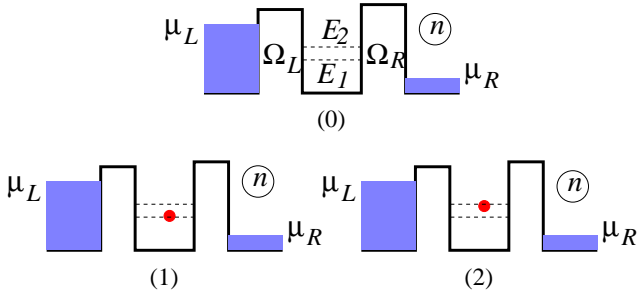,width=3.3in}
 \protect\caption{
   Three available states of the system. Here $n$ denotes the number
  of electrons arriving at the collector by time $t$.  
}
\label{f2}
\end{figure}

Consider again the resonant tunneling through the two levels,
(Fig.\ref{f1}). Let us assume that the Coulomb repulsion of electrons inside
the dot $U_C$ is large such that two electrons cannot
occupy the dot. Then, there are only three available states of the
system, shown in Fig.~\ref{f2}. 

Let us apply (\ref{a5}) by assigning $\alpha,\beta =\{0,1,2\}$,
in an accordance with the states shown in Fig.\ref{f2}. Since the states $1$ and
$2$ are not directly coupled, the corresponding hopping amplitude
$\tilde\Omega=0$ in (\ref{a5}). However, these states can be
connected through the reservoirs [the third and the forth terms of (\ref{a5})]. 
We assume that the corresponding couplings are weakly
dependent on the energy, so that
$|\Omega_{l,r}^{(1)}|=|\Omega_{l,r}^{(2)}|=|\Omega_{L,R}|$.
However, the sign of the $\Omega_{L,R}^{(1)}$ may be the opposite one
with respect to the sign
of $\Omega_{L,R}^{(2)}$. Note that, in the 1-D case, a sign of the product 
$\Omega^{(j )}_L\Omega^{(j )}_R$ is
determined by a sign of the product
$\phi_j (\bar x_l)\phi_j (\bar x_r)$, [see (\ref{a4})]. The latter  
can be positive or negative, depending on the number of nodes of
$\phi_j (x)$ inside the dot\cite{hac}.
Thus, for a 1-D dot, the product
$\Omega^{(j )}_L\Omega^{(j)}_R$ changes its sign when $j\to j+1$.
The reason is that 
the corresponding wave functions $\phi_j(x)$ and $\phi_{j+1}(x)$
differ by an additional node. Hence, the ratio
\begin{equation}
  \eta={\Omega^{(j+1)}_L\Omega^{(j+1)}_R\over \Omega^{(j)}_L\Omega^{(j)}_R}
  \label{aa7}
\end{equation}
is $-1$ for a 1-D dot. However, in the case of 
a three-dimensional (3-D) quantum dot, where the corresponding coupling $\Omega$ is
given by (\ref{a3}), this condition does not hold. 

Taking into account (\ref{aa7}) one obtains from (\ref{a5}) the following
quantum rate equations describing the electron transport through two
levels
\begin{IEEEeqnarray}{lCl}
\label{a8}  
\dot\sigma_{00}^n&=&-2\Gamma_L\sigma_{00}^n+\Gamma_R(\sigma_{11}^{n-1} + 
\sigma_{22}^{n-1})\nonumber\\
&&~~~~~~~~~~~~~~~~~~~~~~~~~
+\eta\Gamma_R(\sigma_{12}^{n-1} + \sigma_{21}^{n-1})
\IEEEyessubnumber\label{a8a}\\
\dot\sigma_{11}^n&=&-\Gamma_R\sigma_{11}^n+\Gamma_L\sigma_{00}^n
-\eta{\Gamma_R\over 2}(\sigma_{12}^n + \sigma_{21}^n)
\IEEEyessubnumber\label{a8b}\\
\dot\sigma_{22}^n&=&-\Gamma_R\sigma_{22}^n+\Gamma_L\sigma_{00}^n
-\eta{\Gamma_R\over 2}(\sigma_{12}^n + \sigma_{21}^n)
\IEEEyessubnumber\label{a8c}\\
\dot\sigma_{12}^n&=&i\epsilon\sigma_{12}^n-\Gamma_R\sigma_{12}^n
+\Gamma_L\sigma_{00}^n-\eta{\Gamma_R\over 2}(\sigma_{11}^n +
\sigma_{22}^n),  
\IEEEyessubnumber\label{a8d}
\end{IEEEeqnarray}
where $\sigma^n_{21}=(\sigma^n_{12})^*$ and 
$\epsilon =E_2-E_1$. In these equations, we assumed that
$\Omega^{(1)}_L=\Omega^{(2)}_L$, so that $\eta
=\Omega^{(1)}_R/\Omega^{(2)}_R$. In the case of a different gauge, 
$\Omega^{(1)}_R=\Omega^{(2)}_R$ and $\eta
=\Omega^{(1)}_L/\Omega^{(2)}_L$, the factor $\eta$ would appear only in
front of the width $\Gamma_L$ in (\ref{a8d}). This of course
does not affect the final result. 

Equations (\ref{a8a})-(\ref{a8d} can be interpreted 
in terms of ``loss'' and ``gain'' terms, and, therefore, they 
represent the quantum rate equations.  
For instance, the first (loss) term in (\ref{a8a})
describes decay of state (0) in Fig.\ref{f2} due to tunneling
of one electron from the left reservoir to the dot. The second (gain) term
of the same equation describes decay of states (1) and (2)
to state (0). The last (gain) term describes decay of the linear
superposition of states (1) and (2).
It is given by the product of the corresponding
hopping amplitudes from the levels $E_{1,2}$ to the collector
reservoir. Since these amplitudes can differ by a sign, this term 
is proportional the relative phase $\eta$ between the states $E_1$
and $E_2$.

It is important to note that all transitions in (\ref{a8})
take place through available continuum states. Therefore, 
the terms $\sigma_{11}^n$ and $\sigma_{22}^n$ in (\ref{a8b}) and 
(\ref{a8c}) can couple with the off-diagonal
matrix elements $\sigma_{12}^n$ through the right reservoir only.
The coupling via the left reservoir would be possible  
for noninteracting electrons through a
new state ($3$) corresponding to two electrons occupying
the levels $E_1$ and $E_2$.
The rate equations in this case would be totally symmetric with
respect to an interchange of $\Gamma_L$ and $\Gamma_R$, and the
result will coincide with that of the single electron description,
[see (\ref{a0}) and (\ref{a1})]. 
Note also that, in the case of $\eta=-1$, the two-level system, shown in
Figs.~\ref{f1},\ref{f2}, can be mapped to a
coupled-dot system. Then (\ref{a8})
turn into the system of quantum rate equations, found earlier for
a description of electron transport through the coupled-dot system\cite{gp,naz}.

On can find that the factor $\eta=\pm 1$, in (\ref{a8}) has the same
meaning as the relative 
phase $\eta$ of two Breit-Wigner amplitudes in Eq.~(\ref{a1}).
Indeed, it is always $-1$ for two subsequent resonances in one
dimensional case. However, in a 3-D quantum dot,
the two subsequent resonances can be found in
the same phase, depending on particular properties of the quantum dot.
One even predicts a whole sequence of the resonances
with the same phases\cite{hac,hac1}. Thus, a measurement of the 
resonance phase $\eta$ could supply us with 
additional information on a quantum dot (impurity) structure,
complementary to spectroscopic measurements. 

Consider first the total current, $I(t)=a I_L(t)+b I_R(t)$, where
$I_{L,R}(t)$, [see (\ref{a6})] are  
the currents in the left or in the right
reservoirs. The coefficients $a$ and $b$ with $a+b=1$
depend on each junction capacitance\cite{but}. For simplicity
we consider only a case where the current in the right reservoir
dominates, $b\gg a$. One easily obtains from (\ref{a8}) that 
\begin{equation}
I(t)=e\Gamma_R[\sigma_{11}(t)+\sigma_{22}(t)
+2\eta\, {\mbox{Re}}\, \sigma_{12}(t)]\, ,
\label{a9}
\end{equation}
where $\sigma_{\alpha\beta}(t)=\sum_n\sigma_{\alpha\beta}^n(t)$. 

Performing summation over $n$ in (\ref{a8}) and solving these
equations in the stationary limit, $t\to\infty$, one easily finds 
for the stationary current $I=I(t\to\infty )$
\begin{equation}
I/e={2\epsilon^2\Gamma_L\Gamma_R\over \epsilon^2\Gamma_R
+2\Gamma_L[\epsilon^2+(1-\eta)\Gamma_R^2]}\, .
\label{a10}
\end{equation}
(Note that the stationary current is independent on the capacitance
of junctions, $a$ and $b$). 

\begin{figure}[htb]
\centering
  \epsfig{file=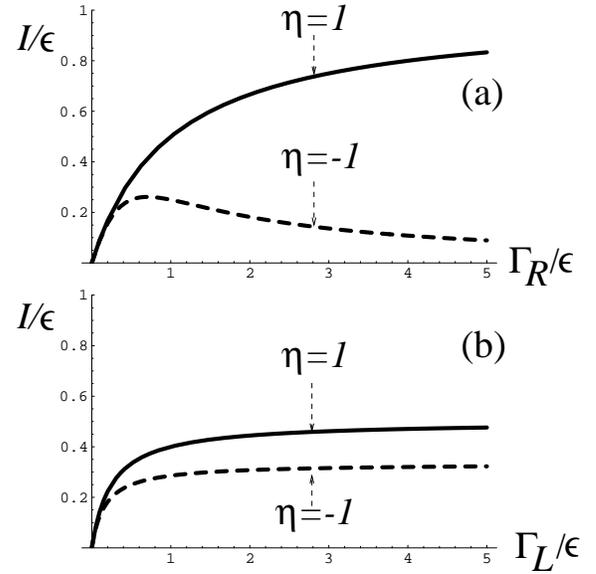,width=3.in}
 \protect\caption{
Total current through two resonance levels (a)   
as a function of the width $\Gamma_R$
for $\Gamma_L =0.5\epsilon$ 
and (b) as a function of the width $\Gamma_L$
for $\Gamma_R =0.5\epsilon$. 
}
\label{f3}
\end{figure}

As expected, when the resonances begin overlap, the current becomes
very sensitive on a sign of the relative phase $\eta$. This
is illustrated Fig.~\ref{f3}(a), where we plot the stationary current $I$ as a
function of the widths $\Gamma_{R}$. One finds that
the current $I$ decreases with $\Gamma_R$ if $\eta =-1$
and increases with $\Gamma_R$ if $\eta =1$.
However, the dependence of the total current $I$ on the
width $\Gamma_L$ [Fig.\ref{f3}(b)] is  rather 
unexpectable. One finds that the current increases for both values of
$\eta$. This is very different from the case
of non-interaction electrons [(\ref{a0}) and (\ref{a1})] where the current is
symmetric under an interchange of $\Gamma_L$ and $\Gamma_R$. Such 
an asymmetry in the case of interacting
electrons is a result of the Coulomb blockade effect\cite{gp,naz}. 
Indeed, an electron enters the dot from the left reservoir with the
rate $2\Gamma_L$. However, it leaves it with the rate $\Gamma_R$,
since the state where the two levels $E_{1,2}$ are occupied
is forbidden due to electron-electron repulsion.
These results can be verified experimentally in the case of a quantum
dot, where the width $\Gamma_{L,R}$ can be varied by changing
the corresponding gate voltage. Then the relative phase
$\eta$ can be obtained from observing the behavior of the resonant
current with $\Gamma_R$ [Fig.~\ref{f3}(a)].

\begin{figure}[htb]
\centering
  \epsfig{file=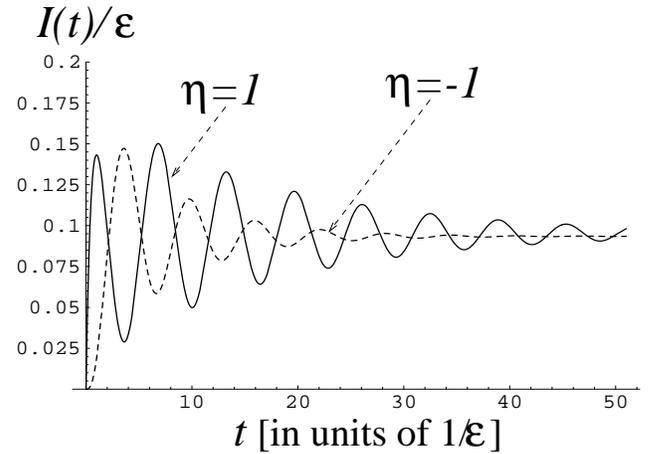,width=3.3in}
  \protect\caption{Time dependence of the resonant current flowing through
    two resonant levels $E_1$ and $E_2$, for
    $\Gamma_L=\epsilon$ and $\Gamma_R=0.1\epsilon$.
    The solid line corresponds to the resonances in phase,  
    $\eta =1$, and the dashed line to the off-phase resonances, $\eta =-1$.
    The dot is empty for $t=0$. 
}
\label{f44}
\end{figure}

The quantum interference effects appear as well in the time-dependent
current. Let us calculate $I(t)$ [(\ref{a9})] by solving
[(\ref{a8})] with the initial conditions
$\sigma_{jj'}(0)=\delta_{j0}\delta_{j'0}$ corresponding to the empty dot.
The time-dependent average current is shown in Fig.~\ref{f44} for
$\Gamma_L=\epsilon$ and $\Gamma_R=0.1\epsilon$ for two values of the
relative phase, $\eta =\pm1$. One finds from this figure that the
current displays strong oscillations in contrast with resonant
tunneling through a single level. The influence of the relative phase
$\eta$ on the quantum oscillations is very substantial. Indeed, the
oscillations related to different values of $\eta$ are shifted
by half of the period and, moreover, the corresponding  
dampings are quite different.

The oscillations in the average current are reflected in the
shot-noise power spectrum given by $S(\omega )=
a S_L(\omega )+ b S_R(\omega )-abS_Q(\omega )$\cite{moz}. Here 
$S_{L,R}$ is the current power spectrum in the left (right) reservoir,
[(\ref{a7})] and $S_Q(\omega )$ is the charge
correlation function of the quantum dot. The latter can also be obtained from
(\ref{a8}). Again, we take for simplicity the case of $b\gg a$, so
that $S(\omega )=S_R(\omega )$. Then one easily finds from
(\ref{a7}) and (\ref{a8}) that 
\begin{equation}
S(\omega) = 2e^2\omega\Gamma_R {\mbox{Im}}\,[Z_{11}(\omega )
+Z_{22}(\omega )+Z_{12}(\omega )+Z_{21}(\omega )]\, ,
\label{a11}
\end{equation}
where
\begin{equation}
Z_{\alpha\beta}(\omega) = \int_0^\infty \sum_n(2n+1)\sigma_{\alpha\beta}^n(t)\exp
(i\omega t)dt\, . 
\label{a12}
\end{equation}
These quantities are obtained directly from (\ref{a8})
by reducing them to the system of linear algebraic equations.

\begin{figure}[htb]
\centering
  \epsfig{file=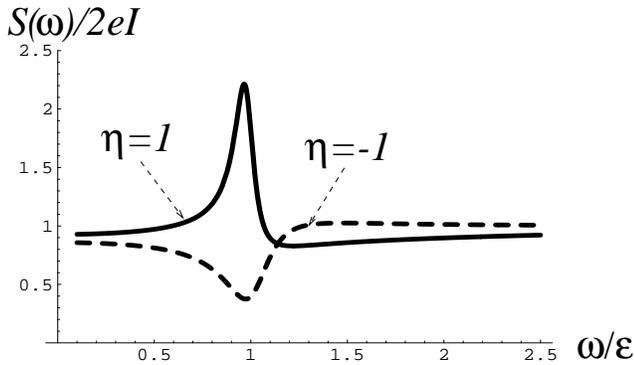,width=3.3in}
 \protect\caption{
   Fano factor versus $\omega$ for the resonant current through two
   levels. The parameters are the same as in Fig.~\ref{f44}.  
}
\label{f4}
\end{figure}

Using (\ref{a11}), we calculate the ratio
of the shot-noise power spectrum to the Schottky noise $S(\omega )/2eI$
(Fano factor), where $I$ is given
by (\ref{a10}). This quantity is shown in Fig.~\ref{f4} for
$\Gamma_L= \epsilon$ and $\Gamma_R=0.1 \epsilon$, which are the same
parameters as in Fig.~\ref{f44}, and 
$\eta =\pm 1$. As expected, the quantum interference is reflected in 
the shot-noise power spectrum. We find that the corresponding Fano
factor shows a peak at $\omega =\epsilon$ in the
case of ``in-phase'' resonances and a dip for out-of-phase resonances.
Although Fig.~\ref{f4} displays the Fano factor 
for an asymmetric quantum dot, $\Gamma_L>\Gamma_R$, such a
strong influence of the phase on the shot-noise power spectrum
pertains in a general case. The effect is merely more pronounced  
for the asymmetric dot. The reason is the Coulomb repulsion that
prevents two electron from occupying the dot
(c.f. with Fig.~\ref{f3}). For noninteracting electrons ($U_C=0$), however,
the effect is mostly pronounced for a symmetric case, $\Gamma_L=\Gamma_R$.

Obviously, the shot-noise spectrum of  resonant current through
a double-dot system should be similar to that shown in Fig.~\ref{f4}
for $\eta =-1$. Indeed, such a system is mapped to a single dot with
two levels, corresponding to the symmetric (nodeless) and
antisymmetric (one-node) states. Therefore the corresponding
shot-noise power spectrum would always show a dip at Rabi frequency
(cf.\cite{bran}),
in contrast with earlier evaluations, which predicted a peak\cite{kor}.

Our results suggest that the measurement of shot-noise spectrum
can be used for a measurement of the relative phase $\eta$.
Technically, it would be more complicated than the
measurement of the total current as a function of
$\Gamma_R$ (Fig.~\ref{f3}), which also determines $\eta$, yet
the measurement of $S(\omega )$ does not distort the dot,
and the phase $\eta$ can be determined  even for
non-overlapping resonances, $\Gamma_{L,R}\ll \epsilon$.

\section{Interference effects in magneto-transport}

\begin{figure}[htb]
\centering
  \epsfig{file=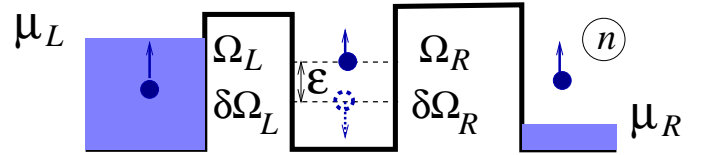,width=3.5in}
 \protect\caption{
    Electron current through an impurity in the presence
    of magnetic field. Here $\epsilon$ denotes Zeeman splitting and 
    $n$ is the number of electrons that have arrived at the right reservoir
(collector) by time $t$.    
}
\label{f5}
\end{figure}

Consider now the electron transport through a quantum dot or 
impurity in the presence of magnetic field. In this case, all of the
levels of the quantum dot are doubled due to the Zeeman splitting
(Fig.~\ref{f5}). Then an electron with spin-up can tunnel only through the upper
level (Fig.~\ref{f5}).
Respectively, an electron with spin-down tunnels only through the lower
level. No interference takes place in this case. However, if 
$g$-factors in the quantum dot and in the reservoirs are different,
the tunneling transitions are accompanied by the spin flip\cite{moz}.
Then the same electron can tunnel from the left to the
right reservoir via two level (cf. with Fig.~\ref{f1}). This process would
generate oscillations in the resonant current in the same
way as was discussed in the previous section.

Let us evaluate the corresponding tunneling amplitudes, which we
denote as $\Omega_{L,R}$ and $\delta\Omega_{L,R}$, respectively, 
for no spin-flip and spin-flip transitions (Fig.~\ref{f5}).
This can be done by using (\ref{a3}). Consider for
the definiteness the electron transitions 
between the dot and the right reservoir. The corresponding
reservoir wave function $\chi_R(\br )$ of (\ref{a3}) is
represented by a Kramers doublet $\chi_R(\bx )=u_R(\br )|\uparrow\rangle +
v_R(\br )|\downarrow\rangle$, where $u_R$ and
$v_R$ are functions of spatial coordinate $\br $ only. Therefore, the tunneling
matrix elements corresponding to the transitions from the resonant
level to the right reservoir without spin flip and accompanied by spin
flip are\cite{moz}
\begin{equation}
  \left (\begin{IEEEeqnarraybox*}[][c]{,c,}
      \Omega_R\\
      \delta\Omega_R
      \end{IEEEeqnarraybox*}\right )
   = -1/(2m) \int_{\br\in\Sigma_R}
   \phi (\br )\stackrel{\leftrightarrow}\nabla_{\bn}
   \left (\begin{IEEEeqnarraybox*}[][c]{,c,}
      u_{R}(\br )\\
      v_{R}(\br )
      \end{IEEEeqnarraybox*}\right )
d {\bsig}
\end{equation}
For relatively small deviations of $g$ factor in the right reservoir
from $2$, $|v| \sim O(|\Delta g u|)$, $\Delta g = g - 2$,\cite{levin},
and so the two transition amplitudes are related as
$|\delta\Omega_{R}| \sim O(|\Delta g \Omega_{R}|)$. For
$\Delta g > 1$, the two components
$u_r$ and $v_r$ are of the same order of
magnitude and so $\delta\Omega_{R}\sim\Omega_{R}$.
The corresponding tunneling amplitudes from the resonant level and the
left reservoir are evaluated in the same way. 

Now we can obtain the quantum rate equations for
magnetotransport through the Zeeman doublet (Fig.~\ref{f5}).
We denote
$\delta\Omega_{L,R}=\alpha_{L,R}\Omega_{L,R}$, where the coefficients
$\alpha_{L,R}$ are of the order of $\Delta g/g$. One finds that, although 
sign[$\delta\Omega_{L,R}$]=$\pm 1$, the product
$\delta\Omega_{L}\delta\Omega_{R}>0$. Thus, the resonances belonging to the Zeeman
doublet are always in phase ($\eta =1$). It is convenient to
write rate equations separately for electrons polarized up and down
in the emitter and collector. Let us consider the polarized up current 
in the emitter and the collector (Fig.~\ref{f5}). Using (\ref{a5}), 
we obtain the following rate equations for the reduced density matrix
$\sigma_{\alpha\beta}^n(t)$ described the spin-polarized transport 
throgh the Zeeman doublet (the index $n$ denotes the number 
of electrons with spin up that have arrived at the right reservoir by time $t$):
\begin{IEEEeqnarray}{lCl}
\label{a13} 
\dot\sigma_{00}^{n}&=&-\Gamma_L(1+\alpha_L^2)\sigma_{00}^{n}\nonumber\\
&+&\Gamma_R(\sigma_{11}^{n-1}+\sigma_{22}^{n})
+\alpha_R^2\Gamma_R(\sigma_{11}^{n}+\sigma_{22}^{n-1})
\nonumber\\
&-&\alpha_R\Gamma_R(\sigma_{12}^{n-1}+\sigma_{21}^{n-1}-
\sigma_{12}^{n}-\sigma_{21}^{n})
\IEEEyessubnumber\label{a13a}\\\
\dot\sigma^n_{11} &=& -\Gamma_R(1+\alpha_R^2)\sigma^n_{11}+\Gamma_L\sigma^n_{00}
\IEEEyessubnumber\label{a13b}\\
\dot\sigma^n_{22}&=&-\Gamma_R(1+\alpha_R^2)\sigma^n_{22}
+\alpha_L^2\Gamma_L\sigma^n_{00}
\IEEEyessubnumber\label{a13c}\\
\dot\sigma^n_{12} &=& i\epsilon \sigma^n_{12}
-\Gamma_R(1+\alpha_R^2)\sigma^n_{12}
-\alpha_L\Gamma_L\sigma^n_{00}
\IEEEyessubnumber\label{a13d}
\end{IEEEeqnarray}
Here we took into account 
that the spin-flip transitions amplitudes, $\delta\Omega$, from the upper 
and lower levels of the quantum dot (Fig.~\ref{f5}) are of the opposite sign.   
Similar to (\ref{a8}) of the previous section, the
quantum interference is generated by transitions between the states
of the Zeeman doublet via the reservoirs. 

Using (\ref{a6}) and (\ref{a13}), one obtains for the spin-up polarized
current in the right reservoir
\begin{equation}
I(t)=\Gamma_R[\sigma_{11}(t)+\sigma_{22}(t)-\alpha_R\sigma_{12}(t)
-\alpha_R\sigma_{21}(t)]\, ,
\label{a14}
\end{equation}
where $\sigma_{\alpha\beta}(t)=\sum_{n}\sigma_{\alpha\beta}^{n}(t)$. 
The corresponding shot-noise power spectrum $S(\omega )$ is given by the McDonald
formula (\ref{a7}). Using (\ref{a13}), we obtain 
\begin{IEEEeqnarray}{lCl}
S(\omega )&=&2e^2\omega\Gamma_R {\mbox{Im}}\, \left \{ Z_{11}(\omega )
+\alpha_R^2Z_{22}(\omega )\right.\nonumber\\
&&\left. ~~~~~~~~~~~~~~~~~~~~-\alpha_R[Z_{12}(\omega )+Z_{21}(\omega
  )] \right \},
\label{a15}
\end{IEEEeqnarray}
where $Z_{\alpha\beta}(\omega )$ is
given by (\ref{a12}). The corresonding Fano factor  
$S(\omega )/2eI$, where $I=I(t\to\infty )$, is therefore determined 
by (\ref{a14}) and (\ref{a15}).

\begin{figure}[htb]
\centering
  \epsfig{file=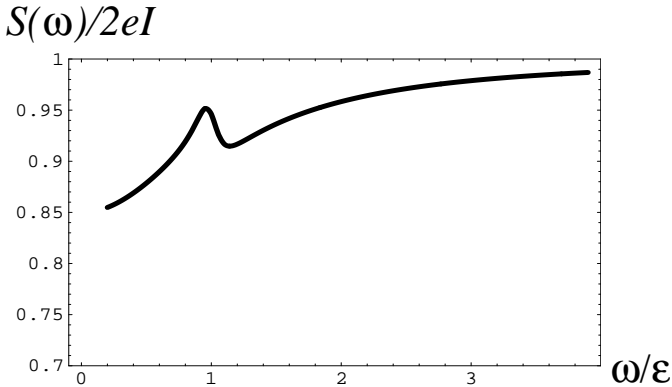,width=3.5in}
 \protect\caption{
   Fano factor versus $\omega$ for the spin-polarized magnetotransport
   current through the Zeeman doublet, shown in Fig.~\ref{f5}.  
}
\label{f6}
\end{figure}

We display in Fig.~\ref{f6} the Fano factor as a function of $\omega$ 
for an asymmetric quantum dot, with the parameters $\Gamma_L=\epsilon$,
$\Gamma_R=0.1\epsilon$ and $\alpha_L=\alpha_R=0.2$. 
This quantity shows a clear peak at frequency close
to the Zeeman splitting\cite{moz}. Similar to the previous case,
discussed in Section II, the effect is 
mostly pronounced for an asymmetric dot due to the influence of 
Coulomb repulsion. Also, we would like to emphasize that  
the two resonances of the Zeeman doublet are ``in phase'', so that $\eta
=1$. Therefore, the shot-noise power spectrum cannot be compared
with that of the current through a couple-dot structure. The latter 
corresponds to $\eta =-1$ and, therefore, the corresponding current
spectrum would always display a dip\cite{bran}, as shown in Fig.~\ref{f4}.

We argued in\cite{moz} that the interference effect in the resonant
tunneling through impurities, considered in the present study, 
can explain coherent oscillations with
Larmor frequency in the STM current. These oscillations were observed
in a set of STM experiments as a peak in the tunneling current
power spectrum\cite{menas,dur}, probably in the spin-polarized
component of the current\cite{dur1}. In fact, there have been
several attempts to explain the experiments\cite{menas},
\cite{molot}-\cite{bul}. 
All these explanations were based on an assumption that the
oscillations of the tunneling current are generated by precession of
a localized spin 1/2, interacting with tunneling electrons.    
In contrast with these models, we 
suggest that it is not the impurity spin but the current itself that 
develops coherent oscillations
due to tunneling of electrons with via the resonant levels of impurity,
spilt  by the magnetic field. Indeed, these oscillation would look
like those generated by a single spin precession, since the Zeeman splitting
coincides with the Larmor frequency. However, there is no precessing
spin in our explanation, but only the interference effect of electrons
moving through two different states\cite{moz}.

An essential requirement for our explanation should be a sizable
spin-orbit coupling effect. This would imply that the $g$-factor
near impurity is different from those inside the bulk and
in the tip. This might be due to low space symmetry of an impurity
on the surface\cite{levi}. Also, the nature of the tip can play a major
role, so that the $g$-factor of the tip would depend strongly on the
tip radius\cite{dur2}.
  
It follows from our arguments that the peak in the STM current spectrum
is not an evidence of a single spin detection, but rather an effect
of coherent resonant scattering (tunneling) on impurity. Nevertheless, the
above described spin-coherence mechanism can be used for a single
nuclear spin detection, as was suggested in\cite{moz}. Indeed, due to
the hyperfine coupling, each electronic level will be split into a
number of sub-levels. Then, according to our model, the peak in STM
current spectrum would be split in a number of peaks corresponding to
transitions between various hyperfine levels. Such a splitting, in fact, has 
already been observed in recent measurements\cite{dur1}. The data 
clearly displays different peaks in the current power spectrum -- evidence
of hyperfine splitting. These experimental results strongly supports our
explanation and opens a new way for a measurement of single nuclear
spin\cite{moz,dur1,dur2}.   

\section{Conclusion}
In this paper, we studied the interference effects in quantum
transport through quantum dots or impurities, where the transport
is carried via several levels. In our investigation, we used a new method
of quantum rate equations which is mostly suitable for treatment
of this type of problems and accounts the Coulomb repulsion in a
simple and precise way. We found that the interference effects
strongly affect the total current as well as the current
power spectrum and depend on the relative phase
of the levels, carrying the current. For instance, 
in the case of out-of-phase resonances, the total current drops down
when the coupling with the collector {\em increases}. This
contraintuitive result represents an 
effect of the destructive interference. On the other hand,
no destructive interference effect would appear when
one increases the coupling with the 
emitter. Such an unexpected asymmetry between the emitter and
the collector does not appear in the case of non-interacting electrons.

We have also demonstrated that the interference effects are reflected in the
shot-noise power spectrum of the resonant current. 
We found that this quantity depends very strongly on the relative phase of the
resonances. It shows a peak for
in-phase resonances and a dip for out-of-phase resonances.
This opens a possibility for studying the internal
structure of quantum dots or impurities by measuring the shot-noise
spectrum of the current flowing through these systems. 

Finally, we applied our method for study the interference effect in
magnetotransport. We showed that, due to the spin-orbit interaction,
the electric current would display the interference effects of
the same nature as in the tunneling through two levels, separated by
the Zeeman spitting. We suggest that this phenomenon can explain the
modulation of STM current found in different experiments and attributed to 
the Larmor precession of the localized spin. Yet, according to our model,
these experiments display the interference effect. The hyperfine
splitting of the signal into several peaks, found in recent
experiments, confirms our model and gives a possibility to use the
interference effect as a new effective tool for a single spin measurements.

\section{Acknowledgment}
The authors is grateful to G. Berman, L. Fedichkin, M. Heiblum, and
D. Mozyrsky for helpful discussions. The author also acknowledges 
very useful correspondence with C. Durcan.

\begin{biography}{Shmuel A. Gurvitz}
received the Ph.D. from the Institute of Theoretical and Experimental
Physics (ITEP), Moscow, Russia, in 1970.

Since 1972, he has been with the Weizmann Institute of Science,
Rehovot, Israel, where he is currently a Professor of Physics
with the Department of Particle Physics. His research interests include
Quantum measurements and decoherence, mesoscopic transport,
multidimensional and cluster tunneling, and deep inelastic scattering.  

\end{biography}

\end{document}